\documentclass[aps,prx,longbibliography,10pt,twocolumn,superscriptaddress,notitlepage,floatfix,amssymb]{revtex4-1}

\usepackage[letterpaper,hmargin={1.5cm,1.5cm},vmargin={1.5cm,2.5cm}]{geometry}

\usepackage[english]{babel}  
\usepackage{mathtools}
\usepackage{float}
\usepackage{amssymb}
\usepackage{amsfonts}
\usepackage{commath}
\usepackage{enumitem}
\usepackage{graphicx}
\usepackage{physics}
\usepackage{braket}
\usepackage{siunitx}
\graphicspath{ {./Figures/} }
\usepackage{color}
\usepackage{longtable}
\usepackage[caption=false]{subfig}
\usepackage[colorlinks,citecolor=red,urlcolor=blue,bookmarks=false,hypertexnames=true]{hyperref}
\usepackage[all]{hypcap}
\usepackage{tikz}
\newcommand*\circled[1]{\tikz[baseline=(char.base)]{
            \node[shape=rectangle,draw,inner sep=2pt] (char) {#1};}}

\makeatletter
\def\@bibdataout@aps{%
\immediate\write\@bibdataout{%
@CONTROL{%
apsrev41Control%
\longbibliography@sw{%
    ,author="08",editor="1",pages="1",title="0",year="1"%
    }{%
    ,author="08",editor="1",pages="1",title="",year="1"%
    }%
  }%
}%
\if@filesw \immediate \write \@auxout {\string \citation {apsrev41Control}}\fi 
}
\makeatother

\begin{document}
\title{Engineering the Level Structure of a Giant Artificial Atom in Waveguide Quantum Electrodynamics}
\author{A.~M.~Vadiraj}
\affiliation{Institute for Quantum Computing and Electrical and Computer Engineering, University of Waterloo, Waterloo, Ontario N2L 3G1, Canada}
\author{Andreas Ask}
\affiliation{Department of Microtechnology and Nanoscience, Chalmers University of Technology, 412 96 Gothenburg, Sweden}
\author{T.~G.~McConkey}
\affiliation{Institute for Quantum Computing and Electrical and Computer Engineering, University of Waterloo, Waterloo, Ontario N2L 3G1, Canada}
\author{I.~Nsanzineza}
\affiliation{Institute for Quantum Computing and Electrical and Computer Engineering, University of Waterloo, Waterloo, Ontario N2L 3G1, Canada}
\author{C.~W.~Sandbo Chang}
\affiliation{Institute for Quantum Computing and Electrical and Computer Engineering, University of Waterloo, Waterloo, Ontario N2L 3G1, Canada}
\author{Anton Frisk Kockum}
\affiliation{Department of Microtechnology and Nanoscience, Chalmers University of Technology, 412 96 Gothenburg, Sweden}
\author{C.~M.~Wilson}
\affiliation{Institute for Quantum Computing and Electrical and Computer Engineering, University of Waterloo, Waterloo, Ontario N2L 3G1, Canada}


\begin{abstract}
Engineering light-matter interactions at the quantum level has been central to the pursuit of quantum optics for decades. Traditionally, this has been done by coupling emitters, typically natural atoms and ions, to quantized electromagnetic fields in optical and microwave cavities. In these systems, the emitter is approximated as an idealized dipole, as its physical size is orders of magnitude smaller than the wavelength of light. Recently, artificial atoms made from superconducting circuits have enabled new frontiers in light-matter coupling, including the study of ``giant" atoms which cannot be approximated as simple dipoles. Here, we explore a new implementation of a giant artificial atom, formed from a transmon qubit coupled to propagating microwaves at multiple points along an open transmission line. The nature of this coupling allows the qubit radiation field to interfere with itself leading to some striking giant-atom effects.
For instance, we observe strong frequency-dependent couplings of the qubit energy levels to the electromagnetic modes of the transmission line. Combined with the ability to \textit{in situ} tune the qubit energy levels, we show that we can modify the relative coupling rates of multiple qubit transitions by more than an order of magnitude. By doing so, we engineer a metastable excited state, allowing us to operate the giant transmon as an effective lambda system where we clearly demonstrate electromagnetically induced transparency. 

\end{abstract}

\date{\today}


\maketitle

\section{Introduction}
Light-matter interaction (LMI) has been one of the most widely explored phenomena in physics. In the microscopic limit, quantum-mechanical interactions can even be engineered between individual photons and atoms. Over the last decades, we have seen many seminal experiments where natural atoms are coupled to quantized electromagnetic fields in high-finesse optical and microwave cavities, a field known as cavity QED~\cite{Thompson:1992jr, Boca:2004gc, Raimond:2001jj, Hyafil:2004ju}. In more recent years, artificial atoms made using Josephson junctions were coupled to superconducting circuits to demonstrate new regimes of light-matter coupling in the microwave domain~\cite{Chiorescu:2004gi, Wallraff:2004dya, FornDiaz:2016bo, Yoshihara:2017bi, Gu:ut, Kockum2019}, a thriving field of research known as circuit QED. 

\begin{figure*}[ht]
\centering
\includegraphics[width=1\linewidth]{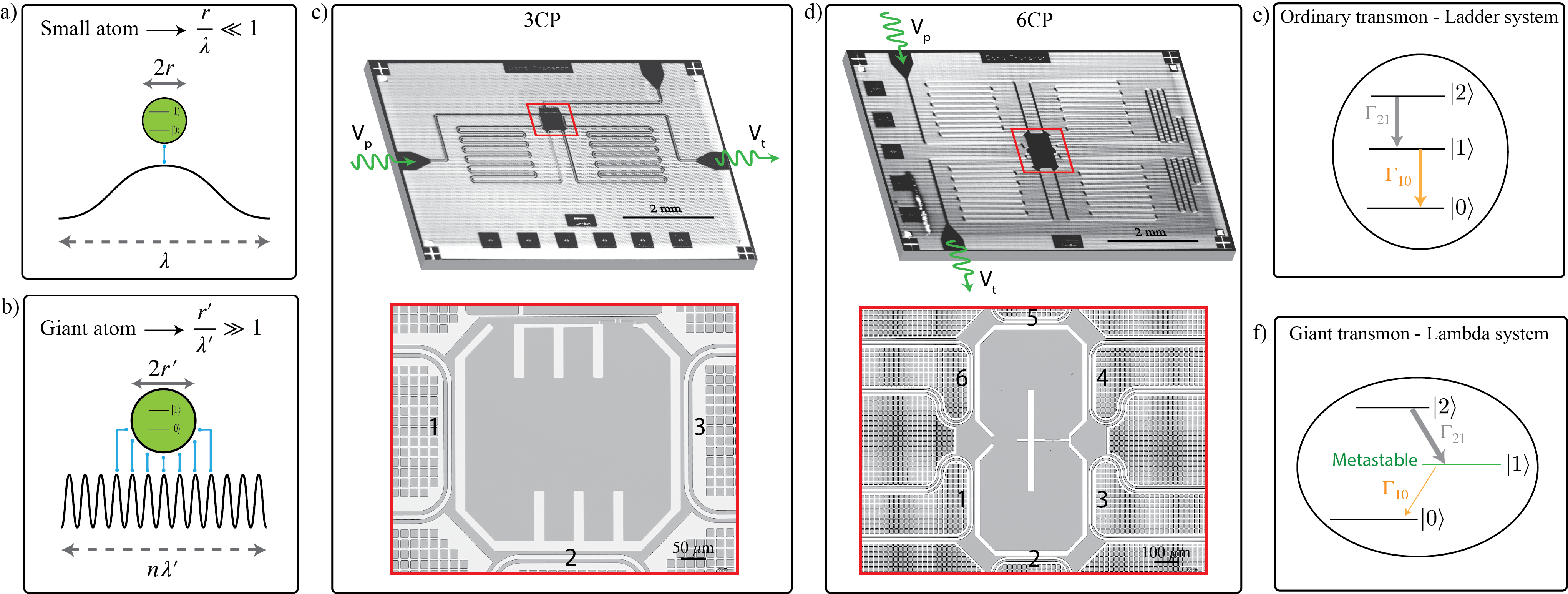}
\caption{Giant artificial atoms in a superconducting waveguide-QED architecture. (a) A conventional atom with a physical size, $r$, much less than the wavelength, $\lambda$, of radiation at its transition frequency. This hierarchy allows the atom (emitter) to be treated as an ideal dipole coupled to the field at a single point. (b) A ``giant" atom with a real or effective size, $r'$, much larger than a wavelength, $\lambda'$. In this work, we produce a giant artificial atom (transmon qubit) with a large effective size by coupling it to the electromagnetic field at multiple points separated by wavelength-scale distances. (c) and (d) show optical micrographs of two giant-transmon devices, with three (3CP) and six (6CP) coupling points respectively, coupled to a long TL, which is suitably meandered. Closeups of the giant transmon structures are also shown for the two devices, with the coupling points and scale bar indicated for clarity. The lighter white regions are aluminum while the darker regions are the silicon substrates. The transmons are designed to have the same $\ket{0}-\ket{1}$ transition frequencies as well as the same anharmonicity between their $\ket{0}-\ket{1}$ and $\ket{1}-\ket{2}$ transitions. (e) Level structure of an ordinary transmon, which has a ladder configuration with $\Gamma_{21}=2\Gamma_{10}$. (f) Level structure of a giant transmon which can be engineered as a lambda system where $\Gamma_{21}\gg\Gamma_{10}$ such that the $\ket{1}$ state can be considered metastable. We use the giant transmons in this configuration to demonstrate electromagnetically-induced transparency (EIT) in both the 3CP and 6CP devices.}
\label{fig:Micrograph_3CP6CP}
\end{figure*}

The simplest theoretical treatments of LMI study the coupling of a two-level emitter, e.g., an atom or qubit, to one or more quantized electromagnetic (EM) modes using a series of approximations. The well-known Jaynes-Cummings model~\cite{JaynesCummings:1963}, which treats the coupling to a single EM mode, uses two key approximations. Firstly, the atom is treated as an ideal dipole, which is a valid approximation when the emitter is much smaller than the wavelength of light, as shown in Fig.~\ref{fig:Micrograph_3CP6CP}(a). Secondly, the dynamics of the LMI are studied under the rotating-wave approximation (RWA), which is applicable when the light-matter coupling strength is still in the perturbative regime. This simplified treatment of LMI has been successful in explaining many phenomena in quantum optics with excellent agreement between theory and experiment~\cite{Gu:ut}. In more complicated systems, where the emitter is coupled to a continuum of electromagnetic modes, the dipole approximation and the RWA are still used with an additional, third assumption that the overall dynamics of the continuum are Markovian. The validity of the Markovian approximation is tied both to the small size of the emitter and the perturbative nature of the coupling.  

Recent developments in microwave quantum optics using superconducting qubits coupled to open transmission lines, a field dubbed waveguide QED, have enabled experiments that demonstrate strong coupling of the qubit to the EM continuum~\cite{Roy:2017hna, Gu:ut, Astafiev:2010cm, Hoi:2013drb, Sathyamoorthy:2014jza, Hoi:2013dh, Hoi:2012jm, Wilson:2011irb}. The ability to tightly confine and guide microwaves in these setups has opened up new avenues for single-photon routing~\cite{Hoi:2011efb}, photon shaping~\cite{FornDiaz:2017hi}, vacuum-mode engineering~\cite{Hoi:2015fha}, interactions between distant qubits with collective decay effects~\cite{vanLoo:2013dfa}, and observation of a large collective Lamb shift~\cite{Wen:2019bs}. Flux qubits have been shown~\cite{FornDiaz:2016bo, Yoshihara:2017bi} to operate in the so-called ultrastrong coupling regime~\cite{Kockum2019}, where the qubit's coupling rate to the EM continuum is comparable to its transition frequency, i.e., far beyond the perturbative regime. As a further consequence of this ultrastrong coupling, the EM spectral density seen by the emitter can no longer be approximated as frequency-independent, making the LMI a non-Markovian process. More recently, non-Markovian phenomena have been explored in a new setting where a transmon qubit is coupled to propagating phonons, in the form of surface acoustic waves (SAWs)~\cite{Andersson:2019cr}. Due to the slow velocity of the waves, the ratio of the size of the transmon to the wavelength of the SAW can be $\sim100$~\cite{Gustafsson:2014gu, Aref:2016cr} [see Fig.~\ref{fig:Micrograph_3CP6CP}(b)]. In this limit, the dipole approximation clearly breaks down and new ``giant atom" effects appear \cite{FriskKockum:2014hd,Kockum:2018hs,Kockum:2019wl,Ask:2019gg}. This includes scenarios where the qubit's radiated field can interfere with itself, creating a variety of non-Markovian behaviors~\cite{Guo:2017ij, Guo:2019ua, ShangjieGuo:2019ua,Ask19}. For instance, nonexponential decay of the qubit was recently observed in such a system~\cite{Andersson:2019cr}.


In this Article, we explore a recent theoretical proposal to realize a giant artificial atom in a waveguide-QED system, where an otherwise conventional transmon qubit is coupled at multiple points to propagating microwaves in a transmission line (TL)~\cite{FriskKockum:2014hd}. The TL is suitably meandered with wavelength-scale distances between the coupling points. Even though the physical size of the transmon is small when compared to the wavelength of interest, the multipoint coupling allows the emission amplitudes of the giant transmon to interfere with themselves, making it an effective giant artificial atom. This interference results in strongly frequency-dependent coupling of its many transitions, an effect that is not seen with an ordinary transmon~\cite{Koch:2007gz}. Stronger modulation of the coupling strengths is made possible by increasing the number of coupling points~\cite{FriskKockum:2014hd}. We present experimental results comparing two separate giant transmons with different numbers of coupling points. We extract the coupling rates of the $\ket{0}-\ket{1}$ and $\ket{1}-\ket{2}$ transitions of the giant transmon as a function of frequency, and show that these can be strongly modulated. We further use this prototype system to engineer the giant transmon into an effective lambda system with a metastable excited state. As a benchmark of the system, we use it to demonstrate a phenomenon characteristic of lambda systems, namely, electromagnetically induced transparency (EIT).

\section{Device}

Our devices each consist of a frequency-tunable transmon qubit capacitively coupled to a one-dimensional (1D) open TL at three (3CP) and six (6CP) coupling points, respectively. The 3CP and 6CP devices were fabricated independently in two separate fabrication runs. Optical micrographs of the two devices are shown in Fig.~\ref{fig:Micrograph_3CP6CP}(c)-(d). The TL, made from aluminum on an intrinsic silicon substrate, is a coplanar waveguide (CPW) with a nominal characteristic impedance of 50~$\Omega$.  The distance between subsequent coupling points for both devices is $\lambda=20.54$~mm. This translates to an intercoupling frequency $\omega_\lambda/2\pi=v/\lambda=5.75$~GHz, where $v=c/\sqrt{\epsilon_{\rm eff}}$ is the phase velocity of microwaves calculated using the effective dielectric constant $\epsilon_{\rm eff}=6.45$ for a CPW on intrinsic silicon.

The giant transmon consists of a superconducting quantum interference device (SQUID) with symmetric Josephson junctions connected to two large electrodes whose capacitance determine the charging energy, $E_c$, of the giant transmon and also provide the coupling to the TL. The geometries of the electrodes are designed such that $E_c$ is similar for both devices. The electrodes are also designed such that the coupling at each point is the same. Q3DExtractor by Ansys was used to design the electrode geometry, targeting $E_c/h\sim450$~MHz. The Josephson junctions of the SQUID were fabricated in aluminum using standard double-angle evaporation. Both devices were designed to have the same Josephson energy, $E_J^{\rm max}$. Additional wirebonds were placed wherever possible between the on-chip CPW ground planes prior to measurement (not shown in Fig.~\ref{fig:Micrograph_3CP6CP}) in order to minimize the effect of parasitic modes. Each device was cooled down separately in a dilution refrigerator with a base temperature reaching $8$~mK. The first transition frequency of the giant transmon, $\omega_{10}/2\pi$, can be tuned from $\sim4-8$~GHz by applying an external magnetic flux, $\Phi$, using a small coil attached to the sample box. Using a heavily attenuated and filtered microwave line, we probe the giant transmon by measuring the transmission coefficient, $t=V_t/V_p$, through the TL. 

Reference~\cite{FriskKockum:2014hd} predicts that the coupling rate of any given transition of the giant transmon to the EM continuum is maximum when the transition frequency is tuned to $\omega_\lambda$. There are also frequency points around $\omega_\lambda$ where this coupling ideally vanishes. Another attractive feature of the giant transmon device is the predicted ability to modulate the relaxation rate of the $\ket{1}-\ket{2}$ transition, $\Gamma_{21}$, relative to that of the $\ket{0}-\ket{1}$ transition, $\Gamma_{10}$. 

This feature opens up new possibilities where interesting three-level physics using a giant transmon can be engineered and explored \cite{FriskKockum:2014hd}. For instance, we can engineer an effective lambda system by having a strongly coupled $\ket{1}-\ket{2}$ transition and weakly coupled $\ket{0}-\ket{1}$ transition. If we can achieve the condition $\Gamma_{21} \gg \Gamma_{10}$, then the $\ket{1}$ level can be viewed as metastable, converting the giant transmon to a lambda system. This is not possible with an ordinary transmon due to its ladder structure where $\Gamma_{21}$ and $\Gamma_{10}$ are of the same order~\cite{Koch:2007gz}. The two situations are described pictorially in Fig.~\ref{fig:Micrograph_3CP6CP}(e)-(f).

\begin{figure}[t]
\center
\includegraphics[width=1\linewidth]{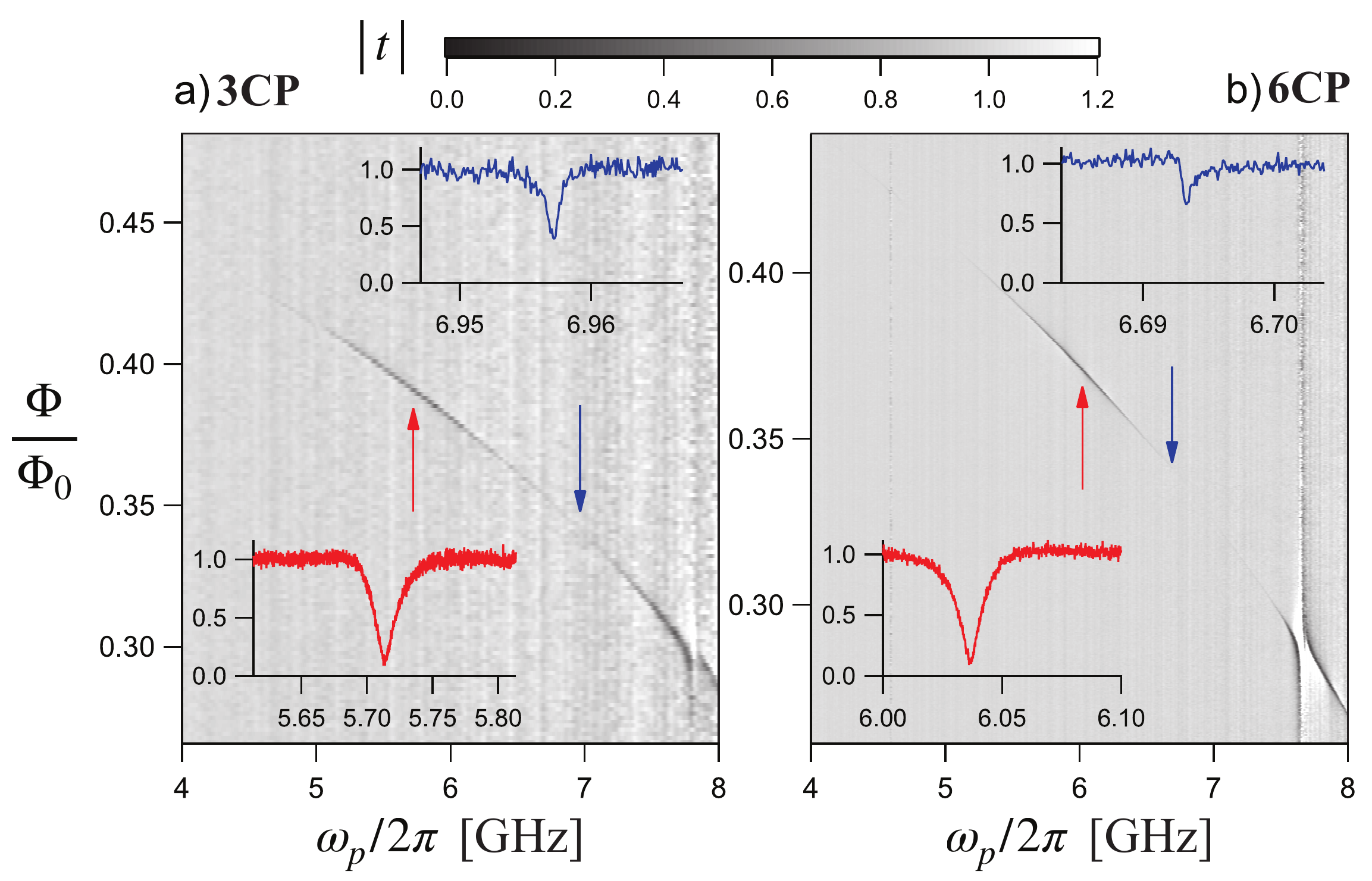}
\caption{Transmission spectroscopy of the giant transmon for the (a) 3CP and (b) 6CP devices for a weak probe at $\omega_p$. We tune the transition frequency of the giant transmon by changing the external magnetic flux, $\Phi$. The color scale indicates the magnitude of the transmission coefficient, $t$. When the transmon is biased close to $\omega_\lambda$ (indicated by red arrows), we see strong extinction of the probe, suggesting that the qubit is strongly coupled to the TL. We also observe frequency regions where the probe's extinction is weak (indicated by blue arrows), implying that the coupling of the transmon to the TL is suppressed. The insets in (a) and (b) are linecuts taken at the flux bias points indicated by the corresponding colored arrows. The background has been subtracted for clarity in both figures.}
\label{WidebandTuningCurves}
\end{figure}

\section{Measurements}
\subsection{$\mathbf{\ket{0}-\ket{1}}$ transition spectroscopy}
\begin{figure}[t]
\center
\includegraphics[width=1\linewidth]{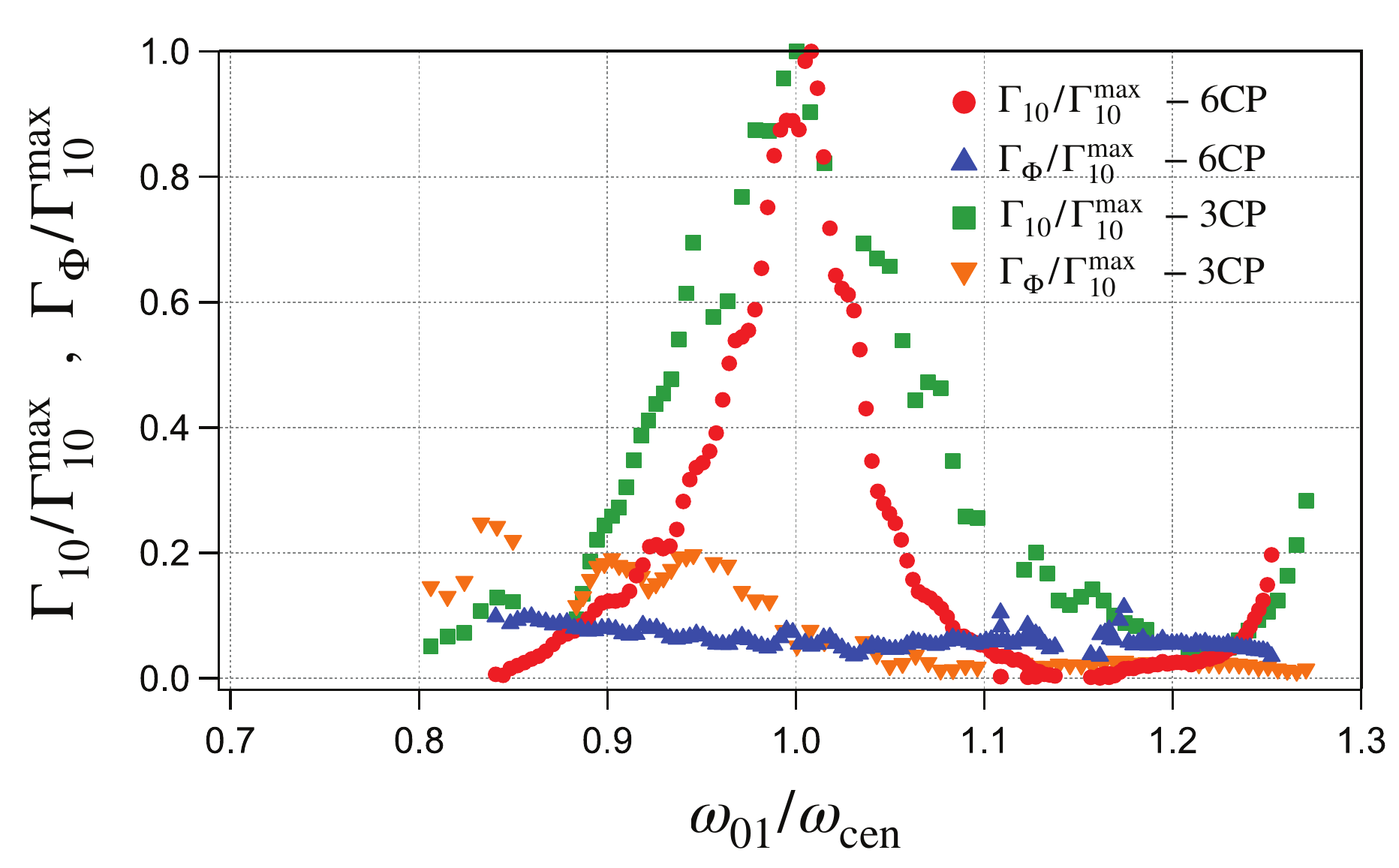}
\caption{Extracted frequency dependencies of $\Gamma_{10}$ and $\Gamma_{\phi}$ for the 3CP and 6CP devices. By fitting transmission spectroscopy data similar to Fig.~\ref{WidebandTuningCurves} using Eq.~(\ref{eqn:TransmissionCoefficient}), we can extract the relevant rates for different qubit frequencies. The rates are normalized to the maximum $\Gamma_{10}^{\rm max}$ (see Table~\ref{tab:Parameters_3CP_6CP}). The transmon frequency $\omega_{10}$ is normalized to the center frequency $\omega_{\rm cen}$ of the rate profiles (extracted from a Lorentzian fit to the profiles). The 6CP device has a sharper relaxation-rate profile, consistent with the theoretical prediction of stronger interference resulting from the larger number of coupling points. The profiles that we extract for the two devices are narrower than the theoretical prediction~\cite{FriskKockum:2014hd} by a factor of approximately 2. However, we see that the FWHM of the 6CP device is approximately half that of the 3CP device, which agrees with the predicted scaling.}
\label{gamma10_gammaPhi_both}
\end{figure}

We first characterize the $\ket{0}-\ket{1}$ transition of the giant transmon by performing transmission spectroscopy. This allows us to extract $\omega_{10}$ and $\Gamma_{10}$ as a function of the external flux bias, $\Phi$. The measured transmission coefficients through the TL for both 3CP and 6CP devices are shown in Fig.~\ref{WidebandTuningCurves}. For a simple qubit in an open TL, the transmission coefficient, $t$, of the probe field can be expressed as \cite{Astafiev:2010cm,Hoi:2011efb}:
\begin{equation}
    t=1-r_{0} \frac{1-\mathrm{i} \delta \omega_{\mathrm{p}} / \gamma_{10}}{1+\left(\delta \omega_{\mathrm{p}} / \gamma_{10}\right)^{2}+\Omega_{\mathrm{p}}^{2} /\left(\Gamma_{10}\gamma_{10}\right)},
\label{eqn:TransmissionCoefficient}
\end{equation}
where $\delta \omega_{\mathrm{p}}$ is the probe frequency detuning from $\omega_{10}$, $\Gamma_\phi$ is the dephasing rate (sum of pure dephasing and non-radiative decay), $\gamma_{10}=\Gamma_{10} / 2+\Gamma_{\phi}$ is the total decoherence rate, $r_o=\Gamma_{10}/2\gamma_{10}$, and $\Omega_{\mathrm{p}}$ is the probe Rabi frequency which is proportional to the probe amplitude. For a weak probe ($\Omega_{\mathrm{p}} \ll \gamma_{10}$) of frequency $\omega_p$, a resonant qubit ($\delta \omega_{\mathrm{p}}=0$) will reflect the incoming probe, resulting in extinction of the transmitted field~\cite{Hoi:2011efb}. The residual transmission is $t=\Gamma_\phi/(\Gamma_\phi+\Gamma_{10}/2)$, such that strong extinction implies $\Gamma_{10}\gg\Gamma_\phi$ and, conversely, weak extinction implies $\Gamma_{10}\ll\Gamma_\phi$.  The strength of extinction hence quantifies the coupling strength of the $\ket{0}-\ket{1}$ transition. In making this statement, we implicitly assume that $\Gamma_\phi$ varies relatively weakly with $\Phi$, which is borne out by the detailed analysis of the experimental data (see Fig.~\ref{gamma10_gammaPhi_both}).

In the spectroscopy data shown in Fig.~\ref{WidebandTuningCurves}, the color scale indicates the magnitude of $t$ after subtracting the background transmission. As expected, when $\delta \omega_{\mathrm{p}}=0$, we see extinction of the transmission amplitude. We also see that the strength of this extinction changes significantly as we vary $\omega_{10}$, indicating a strong frequency-dependent coupling of the $\ket{0}-\ket{1}$ transition. We note that the $\ket{0}-\ket{1}$ transition couples most strongly to the TL near $\omega_\lambda$, as expected from theory. We also see frequency regions far from $\omega_\lambda$ where we observe weak extinction, indicating weak coupling.

\begin{table*}[t]
\caption{Parameters for the 3CP and 6CP devices. All values are expressed in GHz except for $\alpha$, $\beta_{\rm max}$, and $\beta_{\rm min}$, which are dimensionless. The quantity $\alpha$ is the ratio of the maximum to the minimum coupling strength of the $\ket{0}-\ket{1}$ transition. The quantity $\beta$ is the ratio of the $\ket{1}-\ket{2}$ to the $\ket{0}-\ket{1}$ coupling strength. The ratios $\beta_{\rm max}$ and $\beta_{\rm min}$ are measured at two different flux biases, one which maximizes and one which minimizes $\beta$.}
\label{tab:Parameters_3CP_6CP}
\begin{ruledtabular}
\begin{tabular}{c c c c c c c c c c}
Device & $E_J^{\rm max}/h$ & $E_c/h$ & $\omega_{\rm max}/2\pi$ & $\omega_{\rm min}/2\pi$ & $\Gamma_{10}^{\rm max}/2\pi$ & $\Gamma_{10}^{\rm min}/2\pi$ & $\alpha$ & $\beta_{\rm max}$ & $\beta_{\rm min}$  \\ \\[-1em] \cline{1-10}
3CP    & 32.13        & 0.460  & 5.734              & 6.936              & \num{25e-3}                & \num{1.1e-3}       & 23       & 13            & 0.26\\
6CP    & 32.13        & 0.429  & 6.036              & 6.827              & \num{17e-3}                & \num{44e-6}     & 380      & 62      & 0.29
\end{tabular}
\end{ruledtabular}
\end{table*}

To further quantify the modulation of the giant transmon's coupling to its environment, we use Eq.~(\ref{eqn:TransmissionCoefficient}) to fit the transmission spectroscopy data for different values of $\Phi$, extracting $\Gamma_{10}$, $\Gamma_{\phi}$, $\gamma_{10}$, and $\omega_{10}$. We do this under weak probing conditions. Figure \ref{gamma10_gammaPhi_both} shows the extracted $\Gamma_{10}$ and $\Gamma_{\phi}$ as a function of $\omega_{10}$ for the 3CP and 6CP devices using high-resolution transmission spectroscopy data similar to Fig.~\ref{WidebandTuningCurves}. The 6CP device shows a narrower relaxation-rate profile than the 3CP device, consistent with stronger interference effects resulting from the higher number of coupling points. The rates are normalized to the maximum relaxation rate $\Gamma_{10}^{\rm max}$ and the extracted $\omega_{10}$ are normalized to the center frequency $\omega_{\rm cen}$ of the rate profile for the respective device. The various parameters, along with the minimum relaxation rate $\Gamma_{10}^{\rm min}$ measured at $\omega_{\rm min}$ for both devices, are tabulated in Table~\ref{tab:Parameters_3CP_6CP}. Defining an on-off ratio for the coupling of the $\ket{0}-\ket{1}$ transition as $\alpha=\Gamma_{10}^{\rm max}/\Gamma_{10}^{\rm min}$, we observe very strong modulation with $\alpha>300$ for the 6CP device. 

 Although the maximum coupling of the $\ket{0}-\ket{1}$ transition occurs at $\sim\omega_\lambda$, we observe that our minimal-coupling points occur at frequencies different from those predicted by theory~\cite{FriskKockum:2014hd}. Generally, we find that the experimental curves are narrower when compared to the theoretical predictions. However, we observe that the full width at half maximum (FWHM) of the 6CP device, ${\rm FWHM}_{6\rm{CP}}=377\;{\rm MHz}$ is narrower than that of the 3CP device, ${\rm FWHM}_{3\rm{CP}}=687\;{\rm MHz}$, by approximately a factor of 2, which is consistent with the scaling predicted by theory~\cite{FriskKockum:2014hd}. The theoretical FWHM of the 3CP and 6CP devices are $1.78\;{\rm GHz}$ and $853\;{\rm MHz}$, respectively.

To study the cause of this discrepancy in the absolute FWHM, we simulate the microwave transmission of the full-chip layout of the 3CP device using HFSS by Ansys. In these simulations, we model the giant transmon as a classical oscillator \cite{Nigg:2012jj}. To do this, the actual geometry of the transmon electrodes are included in the HFSS simulation, but the SQUID is modelled as a linear inductance. By varying this inductance, we tune the frequency of the oscillator and extract its linewidth around  $\omega_\lambda$. We also take into account the physical configuration of on-chip wirebonds in our simulation. The results of the simulation show a qualitatively similar deviation from theory, but the narrowing is not as large as in our experimental results. If we add more wirebonds to the simulation (beyond what is possible to replicate in experiment), we do, however, find a good agreement with theory. Based on these simulation results, and others discussed in Appendix~\ref{Appendix:HFSS}, we attribute the experimental deviation to parasitic microwave effects such as slot-line modes, radiation effects, etc. We expect that future device designs, for instance, incorporating air bridges between the ground planes, would eliminate these effects.



\begin{figure*}[t]
\center
\includegraphics[width=0.8\textwidth]{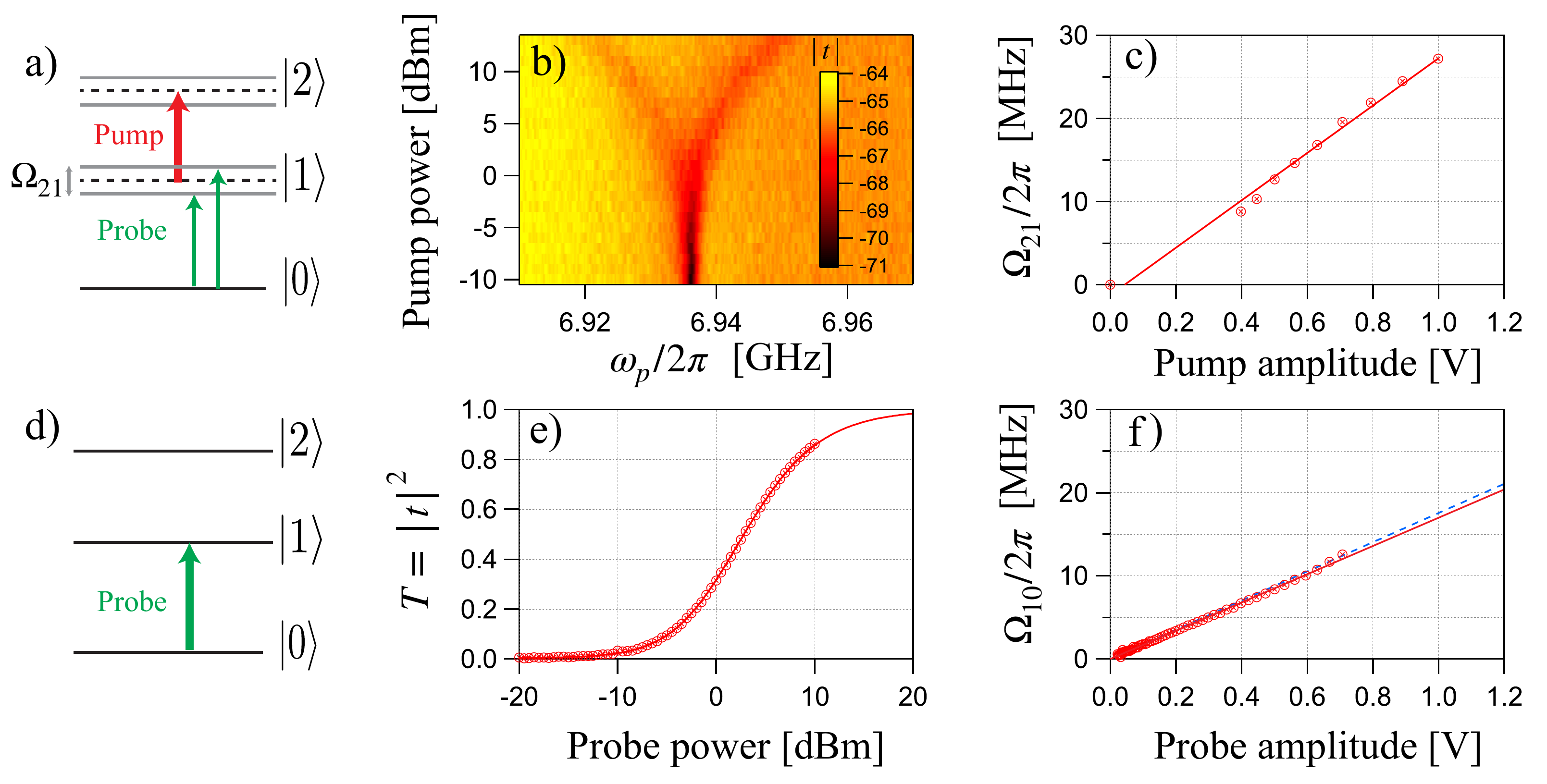}
\caption{Extracting $\Gamma_{21}(\omega)$ by measuring the atom-field coupling constants. (a)-(b) To calibrate the coupling constant $k_{21}$, we use the Autler-Townes splitting (ATS). As illustrated in the level diagram, we observe the ATS by pumping the $\ket{1}-\ket{2}$ transition on resonance and probing the $\ket{0}-\ket{1}$ transition. The pump tone dresses the $\ket{1}-\ket{2}$ transition and we observe the familiar spectroscopic doublet with a splitting given by $\Omega_{21}$. The color scale indicates $|t|$ in dB. (c) We manually extract $\Omega_{21}$ at each power and plot the extracted values (symbols) as a function of pump amplitude, $\sqrt{P}$. Recalling that $\Omega_{ji} =\sqrt{2}k_{ji}\sqrt{P}$ (see text), we extract $k_{21}(\omega)$ from a straight line fit (solid line). (d)-(e) To calibrate $k_{10}$, we use the strong saturation of the transmon's $\ket{0}-\ket{1}$ transition as a function of probe power. This is described by the presence of the probe Rabi frequency $\Omega_p=\Omega_{10}$ in Eq.~(\ref{eqn:TransmissionCoefficient}). After changing the flux bias such that the $\ket{0}-\ket{1}$ transition is at the same frequency as the $\ket{1}-\ket{2}$ transition above, we measure $t$ as a function of the probe power. To characterize the saturation, we plot the transmittance $T=\abs{t}^2$ on resonance, i.e., $\omega_p = \omega_{10}$ (symbols). We fit $T$ using Eq.~(\ref{eqn:TransmissionCoefficient}) (solid line), substituting $\Omega_p=\sqrt{2}k_{10}\sqrt{P}$ and then extract $k_{10}$ as a fitting parameter (solid line). (f) As a second method to extract $k_{10}$, we fit the full transmission curve at each power and extract an independent value of $\Omega_{10}$. The extracted values are plotted versus probe amplitude (symbols). We then extract $k_{10}$ from a straight-line fit to this data (red, solid line). For reference, we also plot the line (blue, dash line) corresponding to the value of $k_{10}$ extracted in panel (e).  There is an obvious agreement between the two values of $k_{10}$. (For subsequent calculations, we use the value of $k_{10}$ extracted from panel (f).)  By using $k_{21}$, $k_{10}$, and $\Gamma_{10}$ (measured independently from low-power spectroscopy), we can infer $\Gamma_{21}(\omega)$ as described in the main text.} 
\label{fig:3CP_RatioMeasurements}
\end{figure*}

\subsection{$\mathbf{\ket{1}-\ket{2}}$ transition spectroscopy}
A novel and interesting aspect of the giant-transmon architecture is the ability to tune the coupling strength of various qubit transitions relative to each other \cite{FriskKockum:2014hd}. Through simple spectroscopy, we have clearly demonstrated the frequency-dependent coupling of the $\ket{0}-\ket{1}$ transition for both the 3CP and 6CP devices. For the $\ket{1}-\ket{2}$ transition, however, extracting the coupling rate $\Gamma_{21}$ is more challenging as we cannot observe this transition directly using single-tone spectroscopy. We can use two-tone spectroscopy to measure the transition frequency $\omega_{21}$, but these measurements do not directly provide $\Gamma_{21}$ \cite{Hoi:2011efb}.

To extract the frequency-dependent $\Gamma_{21}(\omega)$ for the giant transmon, we start by observing that the relaxation rate of the $\ket{i}-\ket{j}$ transition, $\Gamma_{ji}(\omega)$, depends on two quantities: the spectral density of environmental fluctuations, $S(\omega)$, and the atom-field coupling constant, $k_{ji}(\omega)$ \cite{Wendin:2007da}. The specific relation is $\Gamma_{ji}(\omega)=k_{ji}^2(\omega)S(\omega)$.  Since we can directly measure $\Gamma_{10}(\omega)$, we see that we can infer $\Gamma_{21}(\omega)$ by only further measuring the ratio $k_{21}(\omega)/k_{10}(\omega)$.  That is, we see that $\Gamma_{21}(\omega)=[k_{21}^2(\omega)/k_{10}^2(\omega)]\Gamma_{10}(\omega)$. (Here we make the implicit choice to absorb the effects of interference into $k_{ji}$.)

To measure the ratio $k_{21}(\omega)/k_{10}(\omega)$, we note that the Rabi frequency of a driven qubit transition, $\Omega_{ji}(\omega)$, also depends on the same coupling constant. Specifically, $\Omega_{ji}(\omega)=\sqrt{2}k_{ji}(\omega)\sqrt{P}$, where $P$ is the power of the Rabi drive \cite{Hoi:2015fha}. Measuring $\Omega_{ji}$ for the two transitions at the same frequency and drive power allows us to immediately calculate $k_{21}(\omega)/k_{10}(\omega) = \Omega_{21}(\omega)/\Omega_{10}(\omega)$. Note that to measure $\Omega_{21}$ and $\Omega_{10}$ at the same frequency implies measuring them at two different flux bias points, since $\omega_{21} \approx \omega_{10} - E_c/\hbar$.

As shown in Fig.~\ref{fig:3CP_RatioMeasurements}, we take advantage of two well-known physical effects to measure $\Omega_{ji}(\omega)$, both of which have the benefit of being self-calibrating. To measure $\Omega_{21}$, we use the Autler-Townes splitting (ATS) \cite{Autler:1955gb,Hoi:2011efb}. To measure the ATS, we weakly probe around $\omega_{10}$ while strongly pumping at $\omega_{21}$. The strong pump dresses the $\ket{1}$ and $\ket{2}$ levels, leading to an observed splitting of the spectroscopic line around $\omega_{10}$ by an amount $\Omega_{21}$, directly giving us our result. To measure $\Omega_{10}$, we use the qubit saturation as a function of probe power, which can be seen in simple single-tone spectroscopy measurements. This effect is fully described by the presence of the probe Rabi frequency $\Omega_p=\Omega_{10}$ in Eq.~(\ref{eqn:TransmissionCoefficient}). As described in Fig.~\ref{fig:3CP_RatioMeasurements}, we measure both $\Omega_{21}$ and $\Omega_{10}$ at many values of $P$, and use the ratios of the slopes to give a more accurate value of $k_{21}(\omega)/k_{10}(\omega)$. 
In order to quantify the relative modulation of $\Gamma_{21}$ and $\Gamma_{10}$, we define a relaxation-rate ratio $\beta=\Gamma_{21}(\omega_{21})/\Gamma_{10}(\omega_{10})$. For an ordinary transmon, the weakly anharmonic ladder structure of its levels gives $\beta=2$ irrespective of the operating bias frequency. However, our results clearly demonstrate that we can modulate $\beta$ by either enhancing or suppressing $\Gamma_{21}$ relative to $\Gamma_{10}$. Table~\ref{tab:Parameters_3CP_6CP} shows both the maximum and minimum values of $\beta$, which strongly deviate from 2.  We also see that increasing the number of connection points can result in stronger modulation of $\beta$. For the 6CP in particular, we achieve a maximum $\beta$ of 62 and a minimum of 0.29, a modulation by more than a factor of 200. The frequency-dependent relaxation rates of the giant transmon, complemented by the freedom in engineering the rates relative to each other, adds a new flavor to the existing waveguide-QED toolbox. 

\begin{figure*}[t]
\center
\includegraphics[width=0.75\linewidth]{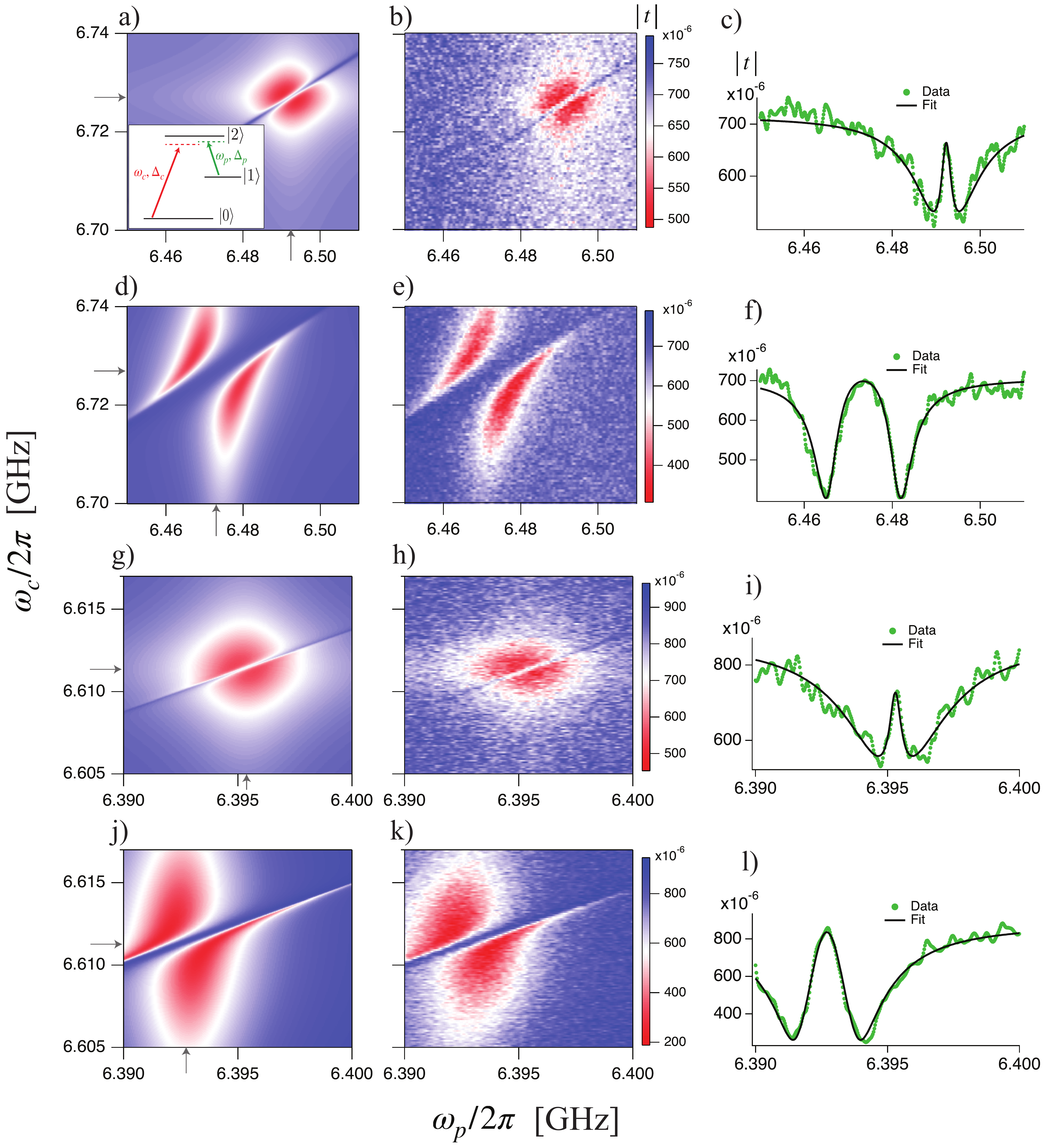}
\caption{EIT in the giant transmon. We study the response of the device in a pump-probe experiment designed to reveal EIT. The inset in (a) shows the three levels of the giant transmon together with the strong pump (control) tone at $\omega_c$ and a weak probe tone at $\omega_p$, with detunings $\Delta_c$ and $\Delta_p$ respectively. (a) and (d) show the numerical calculations for low and high control power, respectively, for the 3CP device. The corresponding experimental data are shown in (b) and (e). The zero-detuning points are indicated by grey arrows on the theory plot axes. The color scale is the magnitude of the transmission coefficient in linear units. The line cuts shown in (c) and (f) are taken at $\Delta_{c}=0$ for the two pump power conditions. The theoretical plots are generated by fitting the measured transmission coefficient using the master equation [Eq.~(\ref{eq:master_eq})]. Where possible, we use independently measured parameters, but the fits also allow us to extract additional parameters (see Table~\ref{tab:Parameters_3CP_6CP}). For the low control power, we are in the EIT regime, while for the high pump power, we are in the ATS regime (see text). The transparency window in the EIT regime appears as a result of destructive interference between the transitions driven by the pump and probe tones. The measurements are done at the flux bias which maximizes $\beta$, such that $\ket{1}$ is metastable. (g)-(l) We repeat similar measurements for the 6CP device where (g)-(i) shows theory and experimental data for low pump power and (j)-(l) for high pump power.}
\label{EIT_2D}
\end{figure*}

\subsection{EIT in a giant transmon}
Three-level atomic physics is a rich branch of AMO, where interference between quantum fields on allowed dipole transitions can lead to interesting physical effects such as the Mollow triplet, ATS, EIT, single-photon lasing, photon blockade, and single-photon transistors~\cite{Gu:ut}. Many of these effects have been demonstrated in circuit-QED setups~\cite{Baur:2009ee,Sillanpaa:2009db,Hoi:2011efb}. 

EIT is a process in which absorption at a given atomic transition is suppressed due to destructive interference between two different excitation pathways in a three-level system \cite{Fleischhauer:2005da}. The suppressed absorption is accompanied by a steep dispersion curve, which can lead to ``slow light" and other phenomena \cite{Hau99,Long:2018kf}. In a three-level lambda system, the interference is enabled by the presence of a metastable state.
A transmon can be used as a three-level artificial atom, but there is an absence of a metastable state, i.e, it cannot be used as a lambda system by itself.

With our giant transmon device, due to its tunable relaxation rates and large modulation of $\beta$, we can engineer a metastable state, turning it into an effective lambda system [see Fig.~\ref{fig:Micrograph_3CP6CP}(f)]. We do this simply by biasing the transmon at the flux that maximizes $\beta$. As a benchmark demonstration of lambda-system physics, we demonstrate EIT in our giant transmon. Our claim to demonstrate EIT, as opposed to ATS, is supported by a detailed master-equation calculation as well as an analysis based on Akaike's information criterion, as suggested in Ref.~\cite{Anisimov:2011fn}. 


The inset in Fig.~\ref{EIT_2D}(a) shows the three levels of the giant transmon and the driving scheme we follow to observe EIT, with the strong pump (control) and weak probe tones. Because the $\ket{0}-\ket{2}$ transition is forbidden as a single-photon process in a transmon, we use a two-photon process, pumping near half the $\ket{0}-\ket{2}$ transition frequency, i.e, $\omega_c = (\omega_{20}-\Delta_c)/2$, where $\Delta_c$ is the pump detuning. The pump transfers population from the ground state $\ket{0}$ to the $\ket{1}-\ket{2}$ manifold. We then apply a weak probe tone at $\omega_p$ to characterize scattering from the $\ket{1}-\ket{2}$ transition.  The probe tone at $\omega_p$ is near $\omega_{21}$ with detuning $\Delta_p = \omega_{21} - \omega_p$. With this driving scheme, the two competing pathways that give rise to EIT are the direct excitation of $\ket{1} \rightarrow \ket{2}$, and the longer path of $\ket{1} \rightarrow \ket{2} \rightarrow \ket{0} \rightarrow \ket{2}$. 

To compare with theory, we calculate the system dynamics by solving the master equation
\begin{equation}
    \begin{split}
        \dot{\rho} &= -i \left[ H,\rho \right ] +\Gamma_{20} \mathcal{D}\left[\sigma_{02}\right]\rho + \Gamma_{21} \mathcal{D}\left[\sigma_{12}\right]\rho \\
         & +\Gamma_{10} \mathcal{D}\left[\sigma_{01}\right]\rho
         + 2\Gamma_{2\phi} \mathcal{D}\left[\sigma_{22}\right]\rho +2\Gamma_{1\phi} \mathcal{D}\left[\sigma_{11}\right]\rho,
    \end{split}
    \label{eq:master_eq}
\end{equation}
where $\Gamma_{ji}$ is the decay rate from state $\ket{j}$ to $\ket{i}$, $\Gamma_{i\phi}$ is the pure dephasing rate of state $\ket{i}$, $\sigma_{ij} = \ketbra{i}{j}$, and we used the notation $\mathcal{D}[X]\rho = X\rho X^{\dagger} - \frac{1}{2}X^{\dagger}X\rho - \frac{1}{2}\rho X^{\dagger}X$ for the Lindblad superoperator \cite{Lindblad76}. The system Hamiltonian, in an appropriate rotating frame (see Appendix~\ref{Appendix:rotating_frame}), is given by
\begin{equation}
\begin{split}
    H & = \Delta_c\sigma_{22} + (\Delta_c - \Delta_p)\sigma_{11} + i\frac{\Omega_c}{2}\left( \sigma_{02} - \sigma_{20} \right) \\
    & + i\frac{\Omega_p}{2}\left( \sigma_{12} - \sigma_{12} \right),
    \end{split}
    \label{eq:Hamiltonian}
\end{equation}
where $\Delta_c = \omega_{20} - \omega_c$ is the detuning of the pump field, $\Delta_p = \omega_{21} - \omega_p$ is the detuning of the probe field, and $\Omega_{c/p}$ is the drive strength of the pump and probe field respectively. Note that we model the $\ket{0} - \ket{2}$ transition as a single-photon process, although the transition is induced via a two-photon process experimentally. To obtain a transmission coefficient, we consider an incoming probe field containing an average number of photons per unit time of $|\alpha|^2$, and use the input-output relation $t = 1 + \sqrt{\Gamma_{21}/2}\braket{\sigma_{12}}{}/\alpha$. To fit the experimental data, $t$ was multiplied by an additional real scale factor to account for amplification and attenuation along the signal line.

There has been discussion in recent literature about how best to distinguish EIT from other phenomena, in particular, the Autler-Townes splitting (ATS) mentioned above~\cite{Anisimov:2011fn}. The question of whether the pump and probe conditions put the system in the ATS or EIT regime can be addressed in a number of ways. From a purely theoretical point of view, the two can be distinguished by examining the poles of the transmission coefficient \cite{Abi-Salloum2010}: $t$ has one pole in the EIT regime and two poles in the ATS regime. The transition between the two regimes can then be parametrized by a threshold pump power, $\Omega_t$, where the number of poles change. By expanding the transmission coefficient to first order in the small parameter $\Omega_p/\Gamma_{21}$, we can derive $\Omega_t = \gamma_{21} - \gamma_{10} = \Gamma_{21}/2 + \Gamma_{20}/2 + \Gamma_{2\phi}$. For $\Omega_c < \Omega_t$ the system is in the EIT regime, and for $\Omega_c > \Omega_t$ the system is in the ATS regime.   

We can transition between the EIT and ATS regime by tuning the pump power. We, therefore, present two sets of transmission measurements for both the 3CP and 6CP device, one with low and one with high pump power. The strength of the pump at the device was extracted from the fits to the master equation calculation, as were the decoherence rates that were not measured independently. Figure~\ref{EIT_2D} compares the measured and calculated EIT curves as a function of $\Delta_c$ and $\Delta_p$, showing very good agreement. The extracted parameters are presented in Table \ref{tab:Parameters_fit}.


\begin{table*}[t]
\caption{Parameters extracted from fitting the transmission coefficient, obtained from the master equation in Eq.~\eqref{eq:master_eq}, to the measured data in Fig.~\ref{EIT_2D}. All values are in units of MHz. Values in squares were extracted independently from other measurements. Parameters marked by an asterisk were not varied during the fitting procedure.}  
\label{tab:Parameters_fit}
\begin{ruledtabular}
\begin{tabular}{l c c c c c c c c r}
Device & $\Gamma_{21}$ & $\Gamma_{20}$ & $\Gamma_{10}$ & $\Gamma_{2\phi}$ & $\Gamma_{1\phi}$ & $\Omega_c$ & $\beta$ & Regime ($\Omega_t$)  \\ \\[-1em] \cline{1-10}
3CP Low power  & 13.6        & 0*  & \circled{1.07}              & 0.94*              & \circled{0.35}                & 3.59       & 12.7       & EIT (7.72)       &     \\
3CP  High power  & 8.92        & 0*  & \circled{1.07}              & 0.94              & \circled{0.35}            & 16.6       & 8.34       & ATS (5.40)       &     \\
6CP Low power    & 2.50        & 0.95  & \circled{0.044}             & 0.67              & 0.11                & 1.03       & 56.8       & EIT (2.40)       &      \\
6CP High power   & 3.93        & 0.06  & \circled{0.044}           & 0.48              & 0.047                & 2.50     & 89.3 & ATS (2.48)       
\end{tabular}
\end{ruledtabular}
\end{table*}

From the decoherence rates in Table \ref{tab:Parameters_fit}, we can calculate the threshold drive-strength $\Omega_t$ and compare it to the extracted drive strength $\Omega_c$ used in the experiment. This suggests, see Table \ref{tab:Parameters_fit}, that the two low-power measurements for both devices are in the EIT regime, the high-power measurement for the 6CP device is just at the border between the two regimes, and the high-power measurement for the 3CP device is in the ATS regime. 

Anisimov et al.~proposed using information-based model selection techniques to distinguish EIT and ATS based on the different predicted absorption profiles for the two processes \cite{Anisimov:2011fn}. The original proposal was to fit two functions: $A_{\text{EIT}} = C_+^2/(\gamma_+^2 + \delta^2) - C_-^2/(\gamma_-^2 + \delta^2)$ and $A_{\text{ATS}} = C^2\left ( 1/(\gamma^2 - (\delta - \delta_0))^2 + 1/(\gamma^2 - (\delta + \delta_0)^2 )\right)$, to the measured absorption spectrum, which is proportional to the real part of the reflection coefficient in our system. We see that the EIT model is formed by the difference of a broad and a narrow Lorentzian centered at the same frequency, whereas the ATS model is the sum of two otherwise identical Lorentzians centered at different frequencies. From the results of the two fits, we then calculate the Akaike information criterion (AIC) for the two models. The AIC is an unbiased estimator of the Kullback-Leibler distance between the proposed model distribution and an (unknown) ``true" model distribution \cite{Burnham2002}. 

For a least-squares fit, the AIC is defined as $I = N\log(\hat{\sigma}^2)+ 2K$ where $K$ is the number of fit parameters, $N$ is the number of data points, and $\hat{\sigma}^2 = \sum \epsilon_i^2/N$ with $\epsilon_i$ being the residuals of the fit. While $I$ for a single model compares it to an unknown true distribution and therefore does not have an easy interpretation, the difference of $I$ for two models has the straightforward meaning of the relative distance of the two models from the true one. In particular, the relative likelihood (probability) of two models is simply given by $\exp(-\Delta_{ij}/2)$ where we define the (positive) Akaike difference $\Delta_{ij} = I_i-I_j$ with $I_i$ the AIC of the i-th model. This relative likelihood is often expressed in a normalized form known as Akaike weights, $w_i$, such that the ratio of the weights $w_i/w_j = \exp(-\Delta_{ij}/2)$ gives the relative likelihood of the two models. 

We calculated the Akaike weights for each measurement in Fig.~\ref{EIT_2D} using the line cuts in (c), (f), (i) and (l) (for fits and details, see Appendix~\ref{Appendix:Akaike}). For the two low-power measurements, Fig.~\ref{EIT_2D} (b) and (h), we find that the EIT model clearly fits better with the relative likelihood of the ATS model being $w_{\text{ATS}}/w_{\text{EIT}} = 10^{-7}$ and $w_{\text{ATS}}/w_{\text{EIT}} = 10^{-8}$, respectively. For the high-power measurement of the 3CP device, with an extracted drive strength well into the ATS regime, the ATS model was strongly favored with the relative likelihood of the EIT model being $w_{\text{EIT}}/w_{\text{ATS}} = 10^{-37}$. For the high-power measurement on the 6CP device, which had an extracted drive strength near the threshold, the EIT model was also strongly favored with $w_{\text{ATS}}/w_{\text{EIT}} = 10^{-30}$. We note that this relative likelihood more strongly favors the EIT model than in the low-power measurements, even though it is near the threshold. Despite this, we can see that the best-fit EIT and ATS curves look qualitatively very similar for this case (Appendix~\ref{Appendix:Akaike}). The fact that the relative likelihoods are much smaller for the two high-power cases may simply be due to the lower signal-to-noise ratio of the low-power measurements.

We believe that the AIC-based testing together with the threshold analysis above convincingly proves that we have observed EIT. 


\section{Conclusions}

In conclusion, we have demonstrated giant artificial atoms in a superconducting waveguide-QED setting. We demonstrated that the giant-atom effects allowed us to modulate the coupling strength of the $\ket{0}-\ket{1}$ transition with an on-off ratio as high as 380. We also showed that we can enhance or suppress the coupling of the $\ket{1}-\ket{2}$ transition relative to the $\ket{0}-\ket{1}$ transition, with a modulation range greater than a factor of 200. This allowed us to engineer the giant transmon into an effective lambda system with a metastable excited state. To further validate this, we clearly demonstrated EIT in our giant transmon, thus benchmarking the quality of our lambda system.  The presence of EIT in our system was verified both by detailed fitting to a master-equation model and by model-selection techniques based on the Akaike information criterion. Our work helps establish giant artificial atoms as a new paradigm in waveguide QED and microwave quantum optics.  

\section{Acknowledgments}
AMV, CWSC, IN, TGM, and CMW acknowledge NSERC of Canada, the Canada First Research Excellence Fund (CFREF), the Canadian Foundation for Innovation, the Ontario Ministry of Research and Innovation, and Industry Canada for financial support. AA acknowledges support from the Swedish Research Council. AFK acknowledges support from the Swedish Research Council (grant number 2019-03696), and from the Knut and Alice Wallenberg Foundation.

\appendix

\section{HFSS simulations}
\label{Appendix:HFSS}

\begin{figure}[h]
\center
\includegraphics[width=1\linewidth]{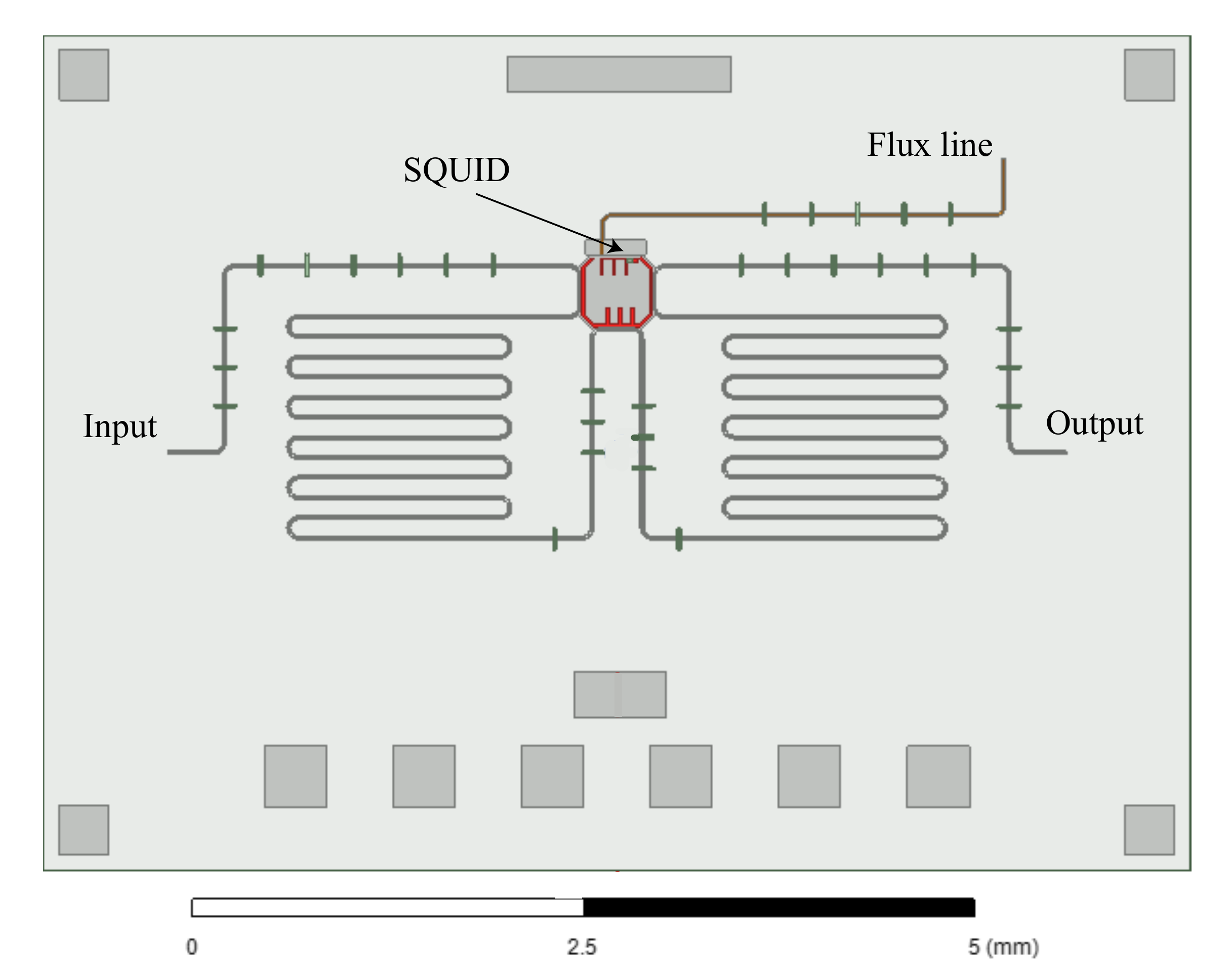}
\caption{3CP device as simulated in HFSS. The wirebonds seen here are representative of those used in the experiments.}
\label{HFSS_3CP_DeviceModel}
\end{figure}

In order to understand the deviation of the experimental relaxation-rate profile from theory (see Fig.~\ref{gamma10_gammaPhi_both}), we performed a full-chip microwave transmission simulation of our 3CP device. We used HFSS from Ansys as our simulation software package. The device model we used for the simulation is shown in Fig.~\ref{HFSS_3CP_DeviceModel}. In the model, we placed wirebonds in the same locations as in the experimental device. Also shown is the DC flux bias line which was not used in the experiment, but was included in the simulations for completeness. 

\begin{figure}[t]
\center
\includegraphics[width=1\linewidth]{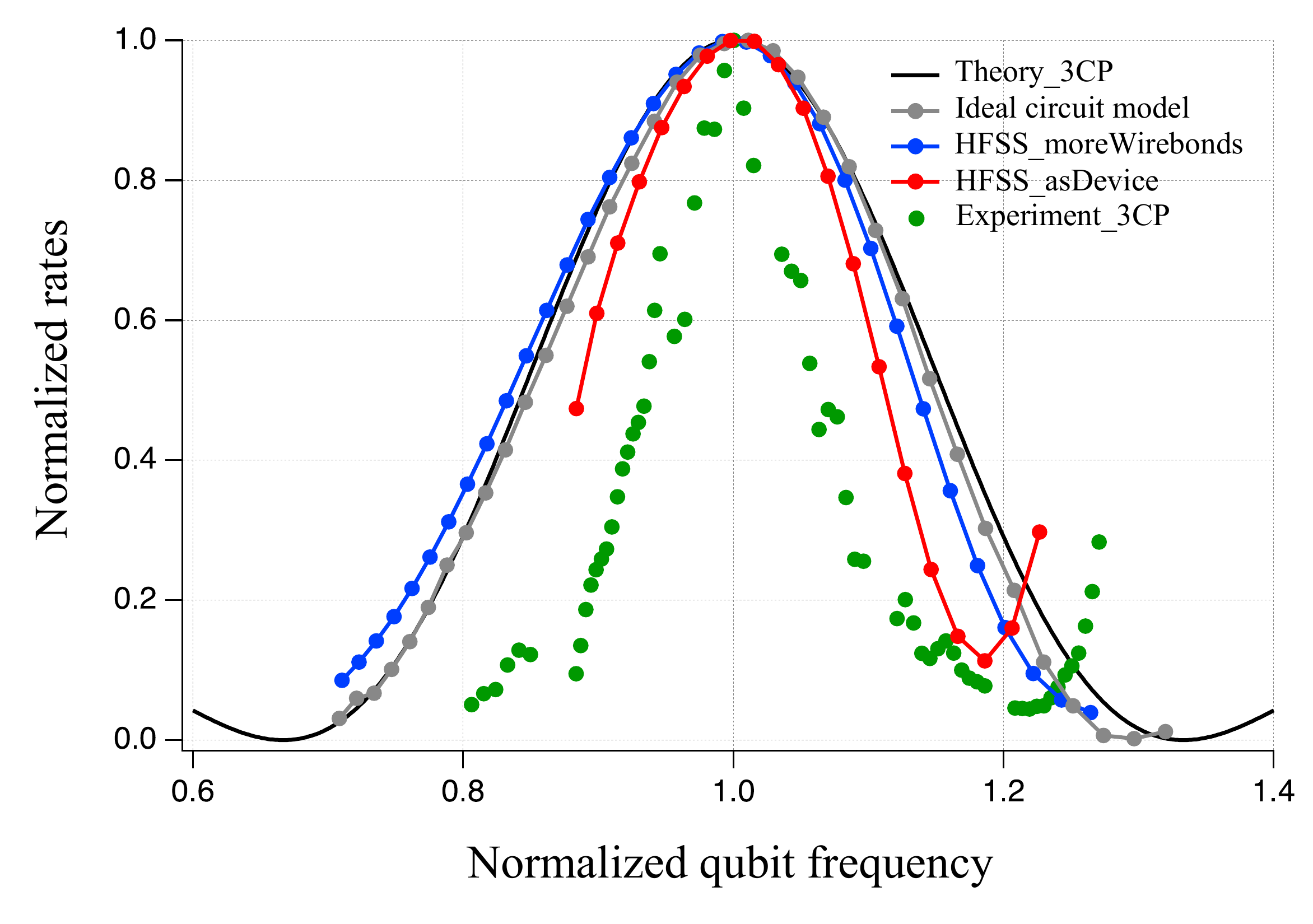}
\caption{Summary of 3CP device simulations. The analytical theory prediction for the 3CP device (black) and experimental data (green) are shown for comparison. We show results of a full-chip, EM simulation done in HFSS, both with a wirebond scheme corresponding to the experimental sample (red) and for the same device with additional wirebonds (blue).  It was not possible to realize the second wirebond scheme experimentally. We also show the results of a simplified closed-form simulation (grey). Both the full EM simulation with extra wirebonds and the closed form simulation agree well with the analytical prediction.  This suggests that the experimental deviation is caused by coupling to stray modes, such as slot-line modes.}
\label{SimulationResults}
\end{figure}

Following the idea of black-box quantization \cite{Nigg:2012jj}, we model the transmon as a classical oscillator which is coupled to the TL at multiple points, as shown in Fig.~\ref{HFSS_3CP_DeviceModel}. As a first simulation, the input and output of the TL, the flux bias line, and the SQUID of the transmon are all modelled as $50\;\Omega$ ports. The physical electrode geometry of the transmon is fully modelled in the simulations (shown in red in Fig.~\ref{HFSS_3CP_DeviceModel}). By carefully designing the mesh around the coupling sections and also the transmon electrode structure, we simulate the S-parameter matrix of the device from $4-8$ GHz. This gives us the background microwave response between the different ports, capturing any parasitic microwave effects of the TL. For the next simulation, we replace the SQUID port with a lumped-element inductor, representing the Josephson inductance of the SQUID. We sweep the value of this inductance across the experimental range, fitting the results of the $S_{21}$ simulation to Eq.~(\ref{eqn:TransmissionCoefficient}) after subtracting the simulated background. In this way, we extract $\Gamma_{10}$ and $\omega_{10}$ for each inductance value. (Note that $S_{21}$ is scattering-matrix notation for the transmission coefficient $t$.) We normalize these rates to the simulated $\Gamma_{10}^{\rm max}$ and normalize $\omega_{10}$ to the center frequency of the rate profile. We plot this in Fig.~\ref{SimulationResults} (red). We see that the simulation profile deviates from theory (black) in a way similar to the experimental data (green), although not to the same extent. 

As a possible explanation for the deviation, we first considered whether we were seeing resonant reflection effects which could arise from the TL impedance discontinuities at the coupling sections of the transmon. We simulated this using a simpler, closed-form microwave model, where the impedance discontinuities were represented as lumped-element inductors in series with the TL. In this case, the transmon is replaced by its equivalent Maxwell capacitance network and the TL sections are constructed using distributed CPW sections. First, we tested the model with the series inductance values set to zero. This produced a rate profile that agrees well with theory. (See the grey curve in Fig.~\ref{SimulationResults}). Increasing the inductance values, thereby introducing reflections in the TL, distorted the rate profile but not in a way that agreed well with the theory.


As a next simulation, we added more wirebonds to the HFSS model in places that were not accessible in the real experiment. The results of this simulation are plotted in Fig.~\ref{SimulationResults} (blue). We observe that this simulation agrees with the theory. This result, along with the result of the idealized simulation above, suggests that parasitic microwave effects which are more pronounced with fewer wirebonds, such as slot-line modes, are responsible for the narrowing of the relaxation-rate profile. In addition, we note that resonant effects that could be caused by the impedance discontinuities in the TL should be captured by the HFSS simulation. However, we would not expect these effects to be suppressed by the addition of more wirebonds. Taken together, these simulations suggest that the performance of future devices could be improved by using a higher-density (microfabricated) solution, such as air bridges, to connect the ground planes of the TL together.

\begin{figure*}[t]
\center
\includegraphics[width=0.85\linewidth]{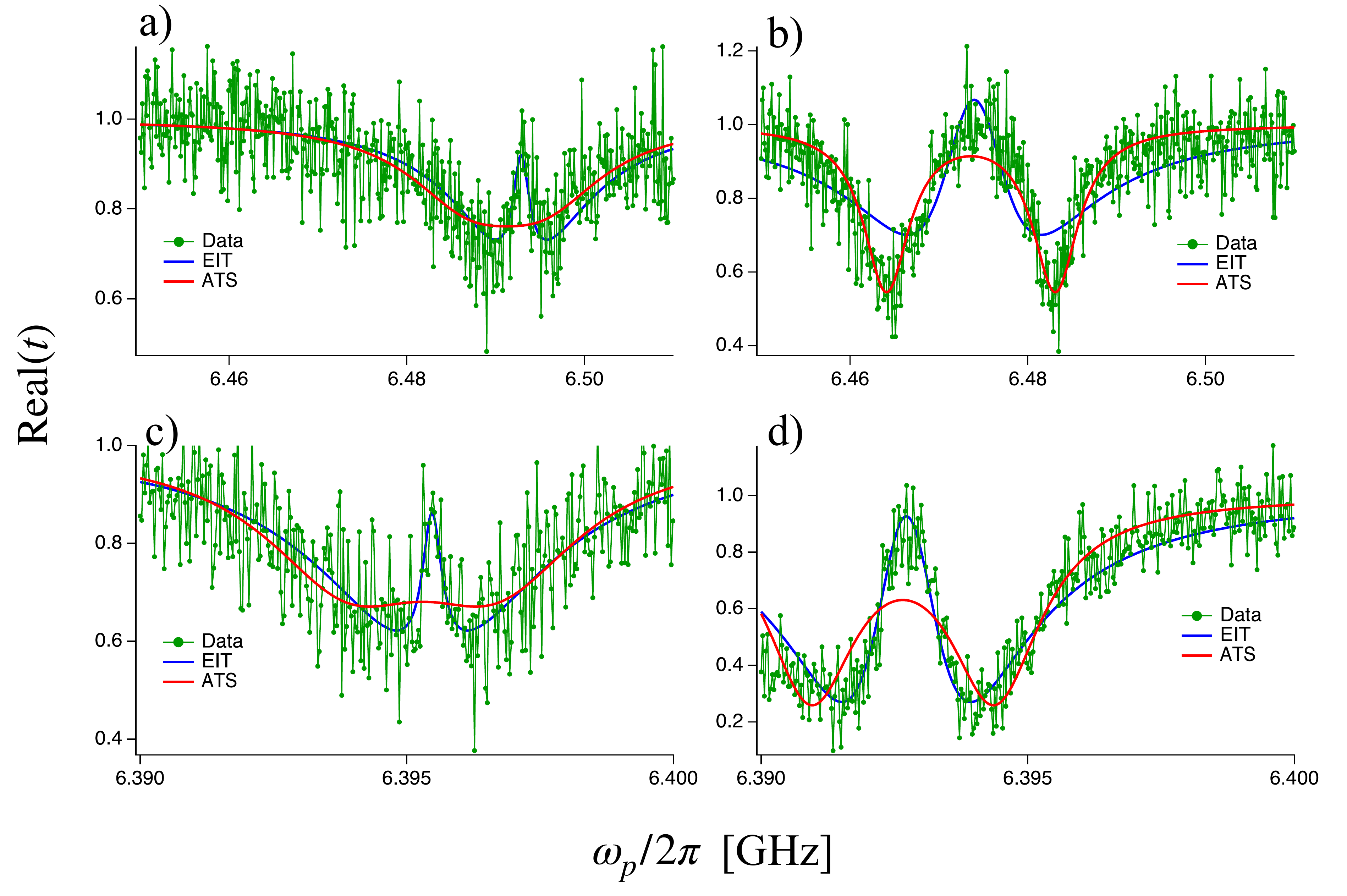}
\caption{Model selection based on Akaike's information criterion. We compare the best fits of an EIT and ATS model to the measured absorption profile, Real$(t)$. In the low-power regime of both (a) the 3CP and (c) the 6CP device, the ATS model does not capture the narrow transparency window which is observed, while the EIT model does. The same is true, although to a lesser extent, for the high-power measurement of the 6CP device shown in (d). This failure is captured in the very small relative likelihood of the ATS model, $<10^{-7}$ in all three of these cases. The ATS model better captures the broad transparency feature observed in the high-power measurement of the 3CP device shown in (b), with a relative likelihood of $10^{-30}$ for the EIT model.}
\label{fig:AkaikeSummary}
\end{figure*}

\section{System Hamiltonian in a rotating frame}
\label{Appendix:rotating_frame}

We consider a three-level system with energy levels $\ket{0},\ket{1}$ and $\ket{2}$, where the decay rate of the $\ket{0} - \ket{1}$ transition is highly suppressed such that a lambda system is formed. The system is driven by a control field at the $\ket{0} - \ket{2}$ transition with amplitude $\Omega_c$ and frequency $\omega_c$, and a weak probe field is applied to the $\ket{1} - \ket{2}$ transition with amplitude $\Omega_p$ and frequency $\omega_p$. The system is described by the following Hamiltonian:
\begin{equation}
\begin{split}
    H &= \omega_{2}\sigma_{22} + \omega_{1}\sigma_{11} + i \frac{\Omega_{c}}{2}\left( e^{-i\omega_ct} + e^{i\omega_ct} \right )\left( \sigma_{02} - \sigma_{20} \right) \\
        & + i\frac{\Omega_{p}}{2}\left( e^{-i\omega_p t} + e^{i\omega_p t} \right ) \left( \sigma_{12} - \sigma_{12} \right),
\end{split}
\end{equation}
where we have set the $\ket{0}$ state to have zero energy, and we have defined $\sigma_{ij} \equiv \ketbra{i}{j}$. The time dependence can be removed by going into a rotating frame by applying the unitary transformation,
\begin{equation}
    U(t) = e^{i t \left( \omega_c \sigma_{22} + (\omega_c - \omega_p)\sigma_{11} \right ) }.
\end{equation}
The rotated Hamiltonian is then given by 
\begin{equation}
    \tilde{H} = UHU^{\dagger} + i\frac{dU}{dt}U^{\dagger}.
\end{equation}
We perform an RWA and neglect terms that oscillate at the frequencies $2\omega_c$ and $2(\omega_c - \omega_p)$, which gives us the time-independent Hamiltonian
\begin{equation}
\begin{split}
    \tilde{H} & = \Delta_1\sigma_{22} + (\Delta_1 - \Delta_2)\sigma_{11} + i\frac{\Omega_c}{2}\left( \sigma_{02} - \sigma_{20} \right) \\
    & + i\frac{\Omega_p}{2}\left( \sigma_{12} - \sigma_{12} \right),
    \end{split}
\end{equation}
where we introduced the two detunings $\Delta_1 = \omega_2 - \omega_p$, and $\Delta_2 = (\omega_2 - \omega_1) - \omega_c$. Note that we have dropped the tilde on the rotated Hamiltonian in Eq.~\eqref{eq:Hamiltonian} of the main text.

\section{Akaike information criterion analysis}
\label{Appendix:Akaike}

In the original proposal to use Akaike's information criterion to distinguish between EIT and ATS \cite{Anisimov:2011fn}, the fitting functions described the absorption spectrum. In this work, the equivalent quantity is the real part of the reflection coefficient. We are not, however, measuring the reflection coefficient directly, but we can still follow the original proposal by using the relation $t = 1 + r$. Before we fit $A_{\text{EIT}}$ and $A_{\text{ATS}}$, mentioned in the main text, to the measured spectra, we do two things: we apply a rotation of the data in order to account for phase shifts induced by propagation delays, and we normalize the data using the amplitude that was extracted from the master-equation simulation. The fits can be seen in Fig.~\ref{fig:AkaikeSummary}. 

\bibliography{GT_Draft_V3.bib}

\end{document}